\documentclass[aps,twocolumn,amsfonts,amssymb,pra,superscriptaddress,10pt]{revtex4-2}

\usepackage{graphicx}
\usepackage{color}
\usepackage{xcolor}
\usepackage{amsfonts}
\usepackage{amsmath}
\usepackage[none]{hyphenat}
\usepackage{subfigure}
\usepackage{orcidlink}
\usepackage[]{hyperref}
\hypersetup{colorlinks=true,citecolor=blue,linkcolor=blue,urlcolor=blue}

\allowdisplaybreaks 

\begin{document}

\title{Odd-parity effect and scale-dependent viscosity in atomic quantum gases} 

\author{Jeff Maki\,\orcidlink{0009-0006-4095-410X}}
\email{jeffrey.maki@uni-konstanz.de}
\affiliation{Department of Physics, University of Konstanz, 78464 Konstanz, Germany}
\affiliation{Pitaevskii BEC Center, CNR-INO and Dipartimento di Fisica, Universit\`a di Trento, I-38123 Trento, Italy}

\author{Ulf Gran\,\orcidlink{0000-0003-0318-0492}}
\email{ulf.gran@chalmers.se}
\affiliation{Department of Physics, Chalmers University of Technology, 41296 Gothenburg, Sweden}

\author{Johannes Hofmann\,\orcidlink{0000-0002-0667-2452}}
\email{johannes.hofmann@physics.gu.se}
\affiliation{Department of Physics, Gothenburg University, 41296 Gothenburg, Sweden}
\affiliation{Nordita, Stockholm University and KTH Royal Institute of Technology, 10691 Stockholm, Sweden}

\date{\today}

\begin{abstract}
Two-dimensional electron gases are predicted to possess an anomalous “tomographic” transport regime that is marked by an odd-even effect in the relaxation times, with odd-parity deformations of the Fermi surface becoming long-lived in comparison to even-parity ones. In this work, we establish that neutral two-component atomic Fermi gases also exhibit this tomographic effect. By diagonalizing the Fermi liquid collision integral, we identify odd-parity modes with anomalously long lifetimes below temperatures \mbox{$T\leq 0.15T_F$}, which is within reach of cold atom experiments. Furthermore, in contrast to electron gases, we find that the odd-even effect in neutral gases is widely tuneable with interactions along the BCS-BEC crossover and is suppressed on the BEC side. We propose as an experimental signature of the odd-even effect the damping rate of quadrupole oscillations, which is anomalously enhanced due to the presence of long-lived odd-parity modes. Our findings suggest that the dynamics of two-dimensional Fermi gases is richer than previously thought and should include additional long-lived modes.
\end{abstract}

\maketitle

\section*{Introduction}

Thermodynamic and transport properties of the normal state of metals as well as ${}^3$He are well described by a semiclassical Fermi gas model~\cite{sommerfeld28,sommerfeld28b}. The microscopic justification of this surprisingly simple description, despite nominally strong interactions, is provided by Landau's theory of the Fermi liquid~\cite{landau57a,landau57b}, which describes thermodynamic and transport properties in terms of a collection of weakly interacting quasiparticles~\cite{pines18,giuliani05,baym04}. The validity of this picture relies on the presence of a Fermi surface and concomitant Pauli blocking, which strongly restricts the available phase space for quasiparticle scattering, leading to well-defined quasiparticles with lifetimes that diverge quadratically at low temperatures, \mbox{$\tau \sim 1/T^2$}. In recent years, Fermi liquid theory has also been applied to a variety of ultracold quantum gas setups, such as minority polarons in strongly spin-imbalanced Fermi gases~\cite{schirotzek09,nascimbene09,Scazza2017, zhenjie19}, the normal phase of balanced Fermi gases above the superfluid transition~\cite{nascimbene10,hu12,vogt12,patel20}, transport and collective modes~\cite{Huang2024} and even few-body systems~\cite{wenz13}. 
For transport measurements, quasiparticle interactions lead to internal dissipative processes with relaxation times that set the magnitude of transport coefficients like the shear viscosity. The general expectation is that these transport times are equal to the quasiparticle lifetimes. 

\begin{figure*}
    \includegraphics{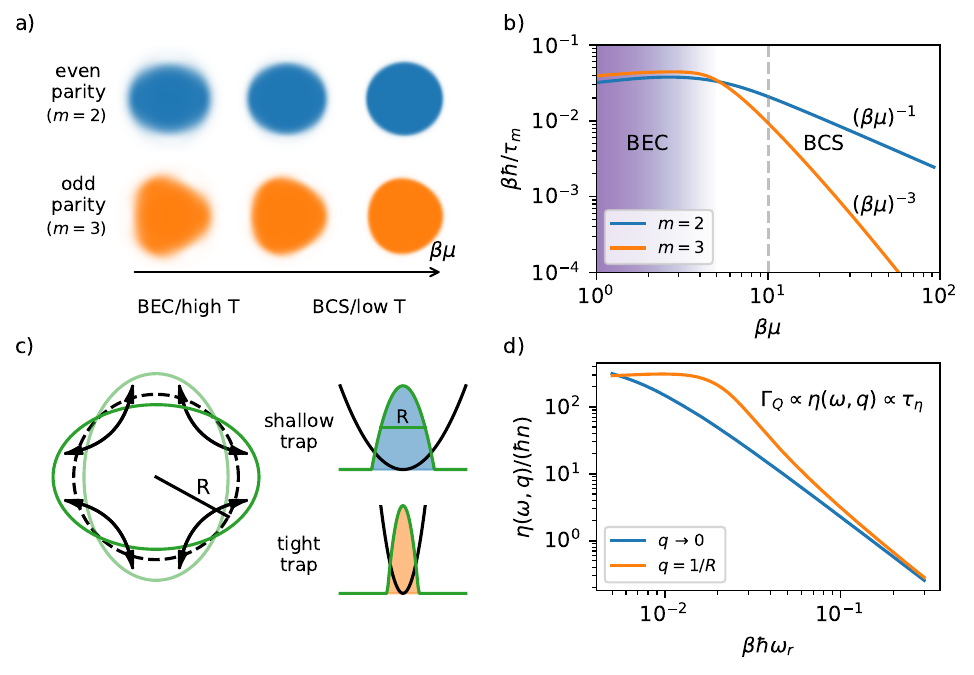}
    \caption{
    \label{fig:relaxation_rates}
    {\bf Tomographic transport in atomic gases.}
    (a)~Nonequilibrium momentum-space distribution for different temperatures $T$ (where $\beta =1/[k_BT]$ is the inverse temperature and $k_B$ is the Boltzmann constant)  and (b)~dimensionless relaxation rates $\beta \hbar/\tau_m$ of the first collective even-parity Fermi surface deformation with  angular momentum~\mbox{$m=2$} (blue) and the first collective odd-parity deformation with~\mbox{$m=3$} (orange). Even (odd) parity refers to an equal (opposite) deformation of the distribution at opposite momenta. With decreasing temperature and increasing density (i.e., increasing \mbox{$\beta\mu$}, where $\mu$ is the chemical potential), the equilibrium Fermi surface becomes sharper and Pauli blocking diminishes the available phase space for scattering, increasing the lifetime of the modes. The conventional Fermi liquid relaxation rate~\mbox{$\sim T^2/\mu$} of the even-parity mode is set by an angular redistribution on the Fermi surface of pairs with equal and opposite momenta. Such a process does not relax the odd-parity modes, however, which have a significantly slower relaxation rate~\mbox{$\sim T^4/\mu^3$}. For interacting Fermi gases along the BCS-BEC crossover, the discrepancy of odd and even-parity lifetimes is present on the BCS side of the crossover, for which the chemical potential is positive, $\beta \mu>0$ (the gray dashed line indicates the non-interacting chemical potential at a moderate small temperature \mbox{$T/T_F = 0.1$} for reference). The effect is then reduced with increasing pairing interaction and vanishes on the BEC side where $\beta \mu<0$. (c)~Real-space dynamics and (d)~damping rate $\Gamma_Q$ as a function of trap frequency $\omega_r$ of the quadrupole mode in a harmonic trap. The quadrupole mode oscillates with frequency \mbox{$\omega_Q = \sqrt{2} \omega_r$} and involves a shear flow in real space. The real-space density profile is shown in the insets and sets the equilibrium size $R$. The damping rate of the quadrupole mode is controlled by the shear viscosity spectral function $\eta(\omega, q)$, which is proportional to the relaxation time of a collective shear deformation of the Fermi surface in momentum space. In a bulk system (shallow trap), this deformation corresponds to the hydrodynamic even-parity mode with \mbox{$m=2$}, but at finite confinement odd-parity modes contribute, and the increased odd-parity lifetime leads to an increased damping rate (orange line) compared to the hydrodynamic prediction (blue line). This damping provides evidence of the enhanced odd-parity lifetime.
    }
\end{figure*}

Recent theoretical work in the context of interaction-dominated electron gases in two dimensions has provided evidence to the contrary and indicates the existence of deformations of the Fermi surface that have  significantly longer lifetimes~\cite{ledwith19,ledwith19b,hofmann23,nilsson24,Fritz2024}: Indeed, a standard Fermi liquid scaling \mbox{$\tau_{\rm e} \sim 1/T^2$} is argued to only apply to a subclass of collective quasiparticle deformations---dubbed even-parity modes---while the remaining modes---dubbed odd-parity modes---decay significantly more slowly as \mbox{$\tau_{\rm o} \sim 1/T^4$}. This leads to the exciting perspective that, at low temperatures, these long-lived modes impact the hydrodynamic description of the gas (in addition to modes that are linked to conserved currents and have infinite lifetime), such that the transport properties of Fermi liquids become richer than commonly assumed. For this reason, there is now intense effort to identify experimental signatures of odd-parity transport, or ``tomographic transport", in electron Fermi liquids~\cite{ledwith19b,hofmann22,hofmann23b,nazaryan24,estrada24,zeng24}. Quite generally, an isolated odd-parity response that is not dominated by even-parity modes appears in transverse probes at finite wavelengths, such as the transverse conductivity~\cite{ledwith19b} or transverse collective modes~\cite{hofmann22}. However, it has so far not been possible to detect such transverse probes in electron Fermi liquids, while in \mbox{${}^3$}He, measurements of transverse sound  have been inconclusive~\cite{roach76,flowers76}. By contrast, quantum gases offer an entirely separate way of inducing transverse dynamics by manipulating the real-space confinement of an external trap. Moreover, since on a microscopic level the odd-even effect is linked to the Fermi statistics and Pauli blocking, one should expect (and as we confirm theoretically) that this effect is not only present for electron gases but also in charge-neutral Fermi liquids. The purpose of this work is thus to initiate a search for the odd-even effect in neutral quantum gases.

The key results of our work, illustrated in Fig.~\ref{fig:relaxation_rates}, are twofold: First, we establish that, just as for electron gases, an enhancement in the lifetime of odd-parity deformations of the Fermi surface also exists for neutral atomic Fermi quantum gases [Fig.~\ref{fig:relaxation_rates}(a)]. From an exact diagonalization of the linearized kinetic collision integral (see methods), we find that these lifetimes differ significantly within temperature ranges accessible in current cold atom experiments [Fig.~\ref{fig:relaxation_rates}(b)]. Moreover, in a marked difference from electron gases, we demonstrate that the effect is widely tunable with interaction strength along the BCS-BEC crossover and vanishes near resonance and on the BEC side of the crossover where the chemical potential $\mu$ is negative and the Fermi surface disappears. As a second key result, we propose as an experimental signature of the odd-even effect the damping rate of collective quadrupole oscillations in a harmonic trap, which involve a shear flow of atoms in real space [Fig.~\ref{fig:relaxation_rates}(c)]. The damping rate of the oscillations is governed by the shear viscosity, which is evaluated at a finite wavenumber set by the trap curvature. As discussed above, this implies that the damping rate is proportional to the relaxation time of the momentum space deformation that is induced by the shear flow in the harmonic trap, \mbox{$\eta \approx np_F v_F
\tau_{\eta}$}. For a shallow trap, the standard hydrodynamic result for the shear viscosity involves an even-parity deformation with conventional Fermi liquid lifetime, 
while for larger confinements, odd-parity modes dominate. Since the shear viscosity is related to the lifetime of a deformation of the Fermi surface induced by the shear flow, the reduced scattering of the odd-parity modes translates to an anomalously enhanced shear viscosity at finite wavenumber and hence an increased damping rate of the quadrupole mode when odd-parity modes become relevant.
Indeed, a full evaluation of the momentum- and frequency-dependent shear viscosity spectral function [Fig.~\ref{fig:relaxation_rates}(d)] shows a strong increase of the damping rate compared to the hydrodynamic prediction for intermediate trap frequencies $\omega_r$ and realistic system parameters. Moreover, as will be discussed below, the effect is enhanced even further both at low temperatures and for small atom numbers, i.e., for mesoscopic Fermi gases. Overall, these findings imply that in experiments with cold gases, the odd parity effect can be widely tuned in a characteristic manner with interaction strength, temperature, harmonic trap frequency, and particle number. Ultracold Fermi gases are thus an ideal platform to study the anomalous relaxation dynamics of collective modes, and it should be readily possible to establish the odd-even effect in the quasiparticle lifetimes by comparing the damping of the quadrupole mode at different experimental parameter values.

\section*{Results}

{\it \bf Quasiparticle lifetimes} We first establish the odd-even effect in the lifetimes of the odd and even parity modes of the quasiparticle distribution function at low temperatures on the BCS side of the normal region of the BCS-BEC crossover (where a well-defined Fermi surface exists) and show that it vanishes along the crossover. Intuitively, this is linked to the fact that fermions along the crossover form increasingly tightly bound dimers, which reduces the chemical potential and destroys the Fermi surface, and with it the Pauli blocking that is at the heart of the odd-even effect. For 2D atomic gases, interactions are short-ranged and described by a delta-function with strength~\mbox{$\hbar^2 \tilde{g}/m^*$} with a dimensionless parameter~\mbox{$\tilde{g}= -2\pi/\ln(k_F a_2)$}, where~\mbox{$k_F$} is the Fermi wavevector, \mbox{$m^*$} is the mass of the atoms and~\mbox{$a_2$} is the 2D scattering length. At a given temperature and density, the interactions interpolate between the BCS side [\mbox{$\tilde{g}<0$} or \mbox{$\ln(k_F a_2)>0$}] and the BEC side [\mbox{$\tilde{g}>0$} or \mbox{$\ln(k_F a_2)<0$}], tuned by either Feshbach ~\cite{Chin10} or confinement-induced resonances~\cite{haller10}.

\begin{figure*}
    \includegraphics{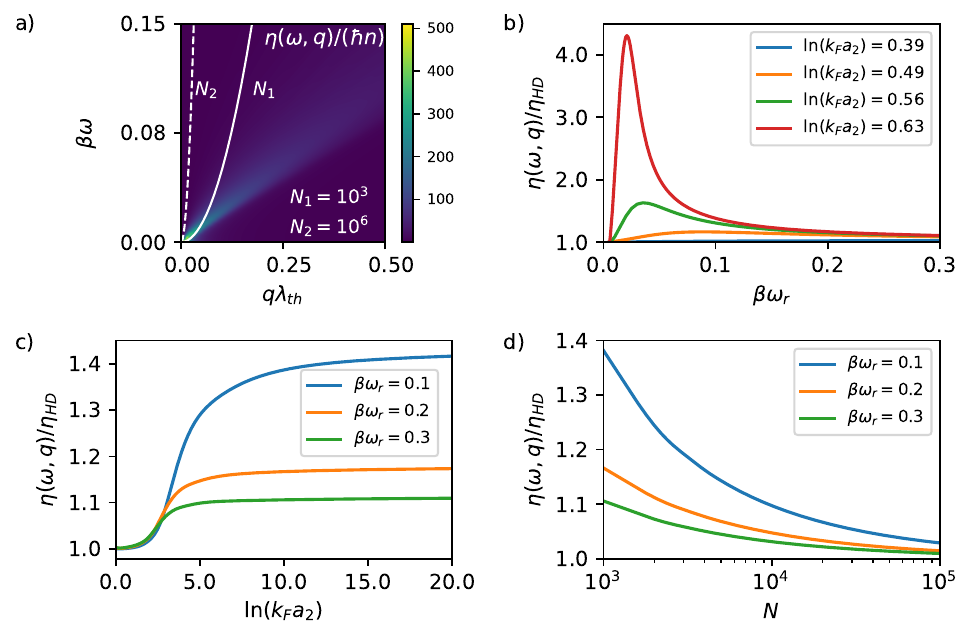}
    \caption{{\bf Shear viscosity spectral function and anomalous damping of the quadrupole mode.} (a) Shear viscosity spectral function in units of $\hbar n$, where $n$ is the particle density at the center of the harmonic trap, as a function of wave number $q$ and frequency $\omega$ (both in thermal units with \mbox{$\lambda_{T} = \sqrt{2\pi \beta\hbar^2/m^*}$}). The white solid (dashed) line describes the path in  parameter space (\mbox{$q = 1/R, \omega = \omega_Q$}) explored by tuning the trap confinement for \mbox{$N_1=10^3$} (\mbox{$N_2=10^6$}) atoms. (b)-(d): Ratio of the shear viscosity spectral function \mbox{$\eta(\omega=\omega_Q,q=1/R)$} and the hydrodynamic shear viscosity \mbox{$\eta_{HD} = \eta(\omega_Q,0)$} as a function of (b)~trap frequency, (c)~interaction strength, and (d)~particle number. Unless otherwise specified, we use an atom number~\mbox{$N=10^3$} and interaction strength~\mbox{$\ln(k_F a_2) = 0.63$} (corresponding to \mbox{$\beta\mu=9$}). In each case, we observe a pronounced deviation from the hydrodynamic prediction for smaller atom number, interactions that are deeper in the BCS limit, and for weaker harmonic traps (or lower temperatures), respectively.}      
    \label{fig:shear_viscosity}
\end{figure*}

The fundamental quantities to describe are the relaxation rates of collective deviations in the quasiparticle distribution from local equilibrium (which is given by a Fermi-Dirac distribution $n_{\rm FD}({\bf p})$). 
These relaxation rates are defined as the eigenvalues of the linearized collision integral (see methods),
\begin{equation}
    \mathcal{L}\left[\psi_m({\bf p})\right] = - \frac{1}{\tau_m} \psi_m({\bf p}) ,
\end{equation}
where the eigenvalue $1/\tau_m$ is the relaxation rate and the eigenfunction \mbox{$\psi_m({\bf p})$} parameterizes a collective quasiparticle deviation as 
\begin{align}
    \delta n({\bf p}) &= n({\bf p}) - n_{\rm FD}({\bf p}) \nonumber \\
    &= n_{\rm FD}({\bf p})(1-n_{\rm FD}({\bf p})) \, \psi({\bf p}) . \label{eq:deltan}
\end{align} 
Due to the rotational invariance of the system, the eigenmodes $\psi_m$ carry a definite angular momentum quantum number \mbox{$m= 0, \pm 1, \pm2...$}. Modes with even (odd) $m$ have even (odd) parity [see Fig.~\ref{fig:relaxation_rates}(a)]. An additional radial quantum number (not indexed explicitly here) describes the number of radial nodes in the eigenfunction, with faster relaxation (i.e., larger $1/\tau_m$) for modes with increasing node number~\cite{nilsson24}. Due to conservation of particle number, energy, and momentum, the lowest eigenvalues with \mbox{$m=0$} and \mbox{$m=\pm1$} are zero. The leading finite even-parity mode has then angular momentum~\mbox{$m=\pm 2$} while the leading odd-parity mode has angular momentum~\mbox{$m=\pm3 $}~\cite{nilsson24}. The corresponding momentum space deformations are shown in Fig.~\ref{fig:relaxation_rates}(a) for three different temperatures. The \mbox{$m=2$} mode sets the dominant contribution to the hydrodynamic shear viscosity~\cite{gran2023shear,frank20,dolgirev24}, while the second mode will give the leading correction in a finite trap geometry and induce the anomalous enhancement of the damping rate.

Figure~\ref{fig:relaxation_rates}(b) shows results for the smallest finite relaxation rates~\mbox{$\tau_m$} with odd and even parity, where we use the expansion method developed in Refs.~\cite{hofmann23,nilsson24} to determine the eigenvalues of the collision integral to arbitrary precision. 
The key observation is the parity-dependence of the lifetime at large \mbox{$\beta \mu$} (\mbox{$\beta = 1/[k_BT]$} and \mbox{$k_B$} is Boltzmann's constant), where Pauli blocking leads to a different scaling of the odd- and even-lifetimes with chemical potential and temperature: odd modes scale as~\mbox{$\beta^4 \mu^3$}, while even modes scale as~\mbox{$\beta^2 \mu$} [indicated in Fig.~\ref{fig:relaxation_rates}(b)]. 
Current experiments reach \mbox{$T/T_F \approx 0.1-0.5$}~\cite{vogt12,Baur2013,Makhalov2014,Boettcher2016,Hueck18}, which is indicated by the dashed gray line in Fig.~\ref{fig:relaxation_rates}(b) for the BCS limit, for which the chemical potential is approximately that of a non-interacting Fermi gas. Already at this point the odd and even lifetimes differ by a factor of two, and even a minor further decrease in temperature (i.e., larger $\beta \mu$) will significantly enhance the odd-even effect: For example, for \mbox{$T/T_F \approx 0.05$} the odd and even lifetimes differ by one order of magnitude. On the BEC side [purple shaded area in Fig.~\ref{fig:relaxation_rates}(b)], the chemical potential becomes negative, \mbox{$\beta \mu <0$}, and scales with the binding energy \mbox{$\epsilon_b= 1/(m^*a_2^2)$} of the two-body bound state,  \mbox{$\mu = -\epsilon_b/2$}. In this regime, odd and even relaxation times are approximately equal and scale as \mbox{$\tau_m \sim e^{-\beta \mu}$}. As discussed, intuitively on the BEC side there is no well-defined Fermi surface and thus one expects the odd-even effect to vanish. This result implies that the odd-even effect is tuned not only by changing the temperature but also the interaction strength. Since the odd-even effect is due to the kinematic constraints of low-energy scattering in the presence of a Fermi surface, it is not overly sensitive to the structure of the two-particle matrix element in the collision integral~\cite{ledwith19, hofmann23}. We have confirmed this by considering various approximations for the two-body scattering matrix, see Supplementary Note I for details.

{\it \bf Shear viscosity} We now propose to observe the odd-even effect along the BCS-BEC crossover by examining the transverse collective dynamics of the gas in real space. In atomic gases, a transverse flow is induced by collective quadrupole modes, which are anisotropic oscillations of a harmonically trapped quantum gas~\cite{Ghosh2002,Bulgac05,Klimin11,Enss2012, Bruun2012,Schaefer2012,vogt12,Chiacchiera2013,Baur2013,Chafin13,hofmann18}. This is shown schematically in Fig.~\ref{fig:relaxation_rates}(c) alongside the radial equilibrium density distribution for the Fermi gas for different trap confinements, which at low temperatures is a Thomas-Fermi profile with radius \mbox{$R= \hbar k_F/(m^* \omega_r) = \sqrt{2\hbar \sqrt{N}/(m^* \omega_r)}$}. The quadrupolar flow is incompressible and there is no change in the internal energy \cite{Ghosh2002,Bulgac05, Klimin11,vogt12}. This yields an oscillation frequency that is independent of the equation of state in the strongly interacting limit,~\mbox{$\omega_Q = \sqrt{2} \omega_r$}~\cite{vogt12}, where \mbox{$\omega_r$} is the frequency of the harmonic trap. In general, the frequency of the quadrupole mode depends on the interaction strength and scales as $\omega_Q = \sqrt{2(2+\tilde{g})/(1-\tilde{g})}\omega_r$ in the collisionless limit~\cite{Ghosh2002}. However, this change in the frequency does not qualitatively affect our analysis, which relies primarily on the finite wavenumber $q$.

The damping rate $\Gamma_Q$ of the quadrupole mode is linked to the local shear viscosity spectral function evaluated at the quadrupole mode frequency, \mbox{$\eta(\omega_Q, {\bf r})$}, via~\cite{Enss2012, Bruun2012,Schaefer2012,vogt12,Chiacchiera2013, Baur2013} 
\begin{align}
    \Gamma_{Q} &= E_{{\rm kin},Q}^{-1} \int d{\bf r} \, \eta(\omega_Q, {\bf r}) 
    \frac{1}{2}\Bigl(\partial_x {\bf v}_Q^y +\partial_y {\bf v}_Q^x\Bigr)^2 . \label{eq:GammaQ}
\end{align}
In this expression, \mbox{$E_{{\rm kin},Q} = \int d{\bf r}\frac{m^*}{2} {\bf v}_Q^2n({\bf r})$} is the kinetic energy associated with the quadrupolar velocity field \mbox{${\bf v}_Q = b (y \hat{\bf x}+x \hat{\bf y})$} (\mbox{$b$} is a constant with units of frequency), and the density of the gas in the harmonic trap~$n({\bf r})$ is evaluated in the local density approximation [see Fig.~\ref{fig:relaxation_rates}(c)]. 
The shear viscosity is a local quantity that depends on the characteristic frequency and wavenumber of the oscillation, with an indirect additional position-dependence (denoted by the subscript) through the local density. 
The wavenumber dependence of the shear viscosity spectral function encodes the odd-even effect and the anomalously long-lived odd-parity modes. Seen in Fourier space, the integral over space in Eq.~\eqref{eq:GammaQ} is equivalent to a trap average that  selects a finite dominant wavenumber $q$ at which the density \mbox{$n({\bf r})$} and current \mbox{$n({\bf r}){\bf v}_Q$} vary during the quadrupole motion. At low temperatures, the density profile is a Thomas-Fermi distribution, and the relevant momentum scale is set by the inverse of the Thomas-Fermi radius: \mbox{$q \sim 1/R$}. The damping rate of the quadrupole mode after trap averaging is then~\mbox{$\Gamma_Q \sim \eta(\omega_Q, q)$}, which in turn is a measure of the relaxation time of shear deformations in momentum space.

In the limit of very shallow harmonic traps, the system is nearly homogeneous and the damping rate is related to the shear viscosity spectral function at~\mbox{$q=0$}. This is the hydrodynamic limit that is dominated by the relaxation of the even-parity~\mbox{$m=\pm2$} mode~\cite{gran2023shear,frank20,dolgirev24}. At finite trap frequencies, i.e., for finite trap curvature \mbox{$q\sim 1/R$}, the odd-parity modes (\mbox{$m=\pm3$} and higher angular momenta) contribute. To obtain an estimate of the characteristic parameter scales at which these odd-parity modes become relevant, we use a relaxation time approximation with a single characteristic even (odd) parity modes decay rate $\gamma_e$ ($\gamma_o$). At low temperatures, these relaxation rates scale as~\mbox{$\gamma_e \sim \tilde{g}^2 T^2/T_F$} and~\mbox{$\gamma_o \sim \tilde{g}^2 T^4/T_F^3$}~\cite{ledwith19b,hofmann23, nilsson24}. When the odd-even effect is present~\mbox{$\gamma_e\gg \gamma_o$}, there is an emergent length scale that defines the tomographic transport limit~\mbox{$\xi = v_F/\sqrt{\gamma_e\gamma_o}$}, where for~\mbox{$q\xi\sim 1$} the system is in the tomographic transport regime, while for~\mbox{$q\xi \ll 1$} is the hydrodynamic regime. In the presence of a harmonic trap, both the relevant wavenumber \mbox{$q \sim \sqrt{\omega_r}/N^{1/4}$} and frequency \mbox{$\omega_Q= \sqrt{2}\omega_r$} are set by the harmonic trap frequency $\omega_r$ and the atom number $N$. Hence, by tuning the trap frequency we explore the shear viscosity spectral function on a line in the~\mbox{$\{\omega,q\}$} parameter space. Expressed in terms of the trap frequency, the condition~\mbox{$q\xi \sim 1$} to observe the odd-even effect in a harmonic trap translates to
\begin{equation}
    q\xi\sim \frac{N \omega_r^3}{\tilde{g}^2 T^3} \sim 1 , \label{eq:oddevencriterion}
\end{equation}
implying that the odd-even effect will be important for~\mbox{$\omega_r \sim \tilde{g}^2 T^3/N$}.

In Fig.~\ref{fig:relaxation_rates}(d), we compare the results for the shear viscosity at finite and zero momentum for~\mbox{$T/T_F = 0.1$} following the Fermi liquid approach outlined in the methods section and Supplementary Note~II. Already for intermediate trap sizes, we see a noticeable difference between the hydrodynamic and finite momentum predictions, which indicates a significant enhancement of the damping rate compared to the hydrodynamic prediction that is due to the anomalous increase in the lifetime of odd-parity modes.

For further illustration, Fig.~\ref{fig:shear_viscosity}(a) shows a spectral plot of the shear viscosity spectral function at finite frequency and momentum as obtained using a Fermi liquid two-body scattering matrix element, where we extract the Landau parameters from experimental data on the low-temperature equation of state~\cite{Boettcher2016}. In general, the shear viscosity spectral function is related to the retarded correlation function of the off-diagonal elements of the stress tensor \cite{Bradlyn2012}, the poles of which give the transverse collective modes. However, in our case the transverse sound pole is diffusive, as the transverse sound mode only becomes real when $F_1>1$ \cite{Khoo2019}. At zero momentum and finite frequency, we find that the spectral function has a Lorentzian shape that describes the standard hydrodynamic-to-collisionless crossover \cite{Enss2012,Bruun2012,Schaefer2012}. 
At finite momentum, the shear viscosity spectral function vanishes at small frequencies, which reflects the fact that the fluid responds instantaneously to the transverse velocity field. At large frequencies the shear viscosity is known to decay as $\omega^{-2}$ in the collisionless limit~\cite{gran2023shear}. Hence one observes a local maximum in the shear viscosity spectral function at finite momentum.
The trajectory \mbox{$(q,\omega)=(1/R,\omega_Q)$} that is traced out in frequency-momentum space  relevant for the damping of the quadrupole mode is indicated by the white lines in Fig.~\ref{fig:shear_viscosity}(a) for two different atom numbers.

Compared to experiments in condensed matter systems, cold atomic gases offer unprecedented tunability of the universal system parameters, such as confinement, interaction strength, and particle number. Figures~\ref{fig:shear_viscosity}(b)-(d) show results for the anomalous enhancement of the shear viscosity for these different parameters evaluated at momentum~\mbox{$q=1/R$} and frequency~\mbox{$\omega=\omega_Q$} compared to the hydrodynamic shear viscosity (i.e., evaluated at zero momentum but finite frequency). First, Fig.~\ref{fig:shear_viscosity}(b) shows the ratio of the shear viscosity as a function of the harmonic trapping frequency \mbox{$\beta \omega_r$} for four interaction strengths \mbox{$\ln(k_F a_2)=0.39,0.49,0.56,$} and $0.63$ (corresponding to chemical potentials \mbox{$\beta\mu =6.8, 7.8, 8.4, 9.0$}, respectively) at fixed atom number~\mbox{$N=10^3$} and temperature \mbox{$T/T_F =0.1$}. On the BCS side of the crossover (i.e., for \mbox{$\ln(k_F a_2)>0$}), there is a strong enhancement of the shear viscosity over the hydrodynamic result at small frequencies. 
As expected we observe hydrodynamic transport in the strongly interacting limit (\mbox{$\ln(k_F a_2) \approx 0$}) implying there is no odd-even effect. This trend continues on the BEC side (\mbox{$\ln(k_F a_2) < 0$}) where the odd- and even-parity relaxation rates are of the same order of magnitude. This can be seen in Fig.~\ref{fig:shear_viscosity}(c), which shows the shear viscosity spectral function as a function interaction strength for three different values of the trap frequency $\beta\omega_r=0.1,0.2,$ and $0.3$, again with \mbox{$N=10^3$} particles and temperature \mbox{$T/T_F = 0.1$}. On the BCS side of the crossover and for finite confinement, 
the shear viscosity is enhanced due to the odd-even effect. The enhancement increases with increasing confinement and reduces to the hydrodynamic result for weak confinement. Finally, in Fig.~\ref{fig:shear_viscosity}(d) we examine the shear viscosity as a function of atom number at fixed interaction strength \mbox{$\ln(k_F a_2) = 0.63$} at \mbox{$T/T_F = 0.1$} for three values of the trap frequency $\beta\omega_r=0.1,0.2,$ and $0.3$. Since a decreasing particle number $N$ increases the momentum as~\mbox{$q=1/R \sim N^{-1/4}$}, we observe a strong enhancement for mesoscopic particle ensembles and a trend toward the hydrodynamic limit for large particle number.

Comparing these results with the criterion in Eq.~\eqref{eq:oddevencriterion}, we indeed find that it accurately predicts the onset of the tomographic regime. As an example, for parameters where we observe an anomalous shear viscosity---such as for \mbox{$\beta \omega_r = 0.1$}, \mbox{$\ln(k_F a) \approx 10$} and \mbox{$N= 10^3$}---we find~\mbox{$q\xi \approx 3.2$}. By contrast, near resonance with~\mbox{$\ln(k_F a) \approx 1$} with the same \mbox{$\beta \omega_r$} and $N$, we find \mbox{$q\xi \approx 0.05$}. 

\section*{Discussion}

We propose neutral quantum gases as a platform to observe and control the odd-even effect in quasiparticle lifetimes, which is an essential but to date unobserved fundamental aspect of Fermi liquid theory. Compared to electron Fermi liquids, neutral quantum gases offer three distinct advantages that will favor an experimental observation of the odd-even parity effect: First, the effect is widely tunable with interaction strength and vanishes on the BEC side of the crossover because the Fermi surface vanishes and Pauli blocking no longer provides a phase-space constraint for scattering. Second, the real-space confinement allows to directly access oscillatory transverse dynamics in real space and probe the shear viscosity through the damping of collective modes. While we here discuss the quadrupole mode damping in detail, in principle, one could also excite other collective modes, like the breathing mode, which is isotropic and will not depend sensitively on the odd-even effect, as a benchmark to examine the anomalous behaviour of the shear viscosity. Third, the odd-even effect depends sensitively on both trap confinement and particle number and is strongly enhanced for mesoscopic Fermi gases. Although the odd-even effect will be present for any particle number as long as the temperature is sufficiently small, our findings suggest that the onset temperature of this effect increases with decreasing particle number, providing an additional way to access this effect. Already a number of experiments have been able to create mesoscopic ensembles of fermions~\cite{peppler2018,Sobirey2022, Brandstetter2023,verstraten2024} and would make an excellent platform for investigating this odd-even effect and its effect on the damping of the quadrupole mode.

\section*{Methods}

We model the dynamics of the two-dimensional Fermi gas in the normal state using Landau's Fermi liquid theory, which describes the non-equilibrium dynamics using a kinetic equation for the fermionic quasiparticles:
\begin{equation}
    \left[\partial_t + \frac{1}{\hbar}\nabla_{\bf p}\epsilon_{\bf p} \cdot \nabla_{\bf r} +\frac{1}{\hbar}\left(\mathcal{F}- \nabla_{\bf r} \epsilon_{\bf p} \right)\cdot\nabla_{\bf p}\right]n({\bf p}) = \mathcal{I}\left[n({\bf p})\right].
\label{eq:FL_Boltzmann}
\end{equation}
Here, \mbox{$n({\bf p}) = n(t,{\bf x},{\bf p})$} is the local distribution function of quasiparticles at time~$t$ and position~${\bf r}$ with momentum~${\bf p}$ (we omit the space-time dependence in the following).  The quasiparticle dispersion in Eq.~\eqref{eq:FL_Boltzmann} is
\begin{equation}
    \epsilon_{\bf p} = \frac{\hbar^2 p^2}{2m^*}+ \int \frac{d{\bf p'}}{(2\pi)^2} f_{\bf p, p'} \delta n({\bf p')} ,
\end{equation}
which depends on the single-particle dispersion with atomic mass $m^*$ and a term that describes the mutual interaction energy from other quasiparticles. The interactions between quasiparticles in this term are parameterized by \mbox{$f_{\bf p,p'}$} and depend on \mbox{$\delta n({\bf p})$}, the deviation of the quasiparticle distribution function away from global equilibrium. For a circular 2D Fermi surface, we expand the interaction in angular harmonics and rescale to a dimensionless form as \mbox{$F_{\bf p, p'} = \nu f_{\bf p, p'} = \sum_{m=-\infty}^{\infty} F_m e^{i m (\theta_{\bf p}-\theta_{\bf p'})}$}, with \mbox{$\nu = m^*/(\pi\hbar^2)$} the density of states at the Fermi surface, $\theta_{\bf p}$ the polar angle of the vector~${\bf p}$ in momentum space, and~\mbox{$F_m$} the dimensionless interaction strength of the \mbox{$m$}-th harmonic. To determine the shear viscosity, we also include an external force ${\bf \mathcal{F}}$ in Eq.~\eqref{eq:FL_Boltzmann} induced by a gradient of a transverse velocity field,
\begin{equation}
{\bf \mathcal{F}}^i = m^*{\bf v_p}^j \nabla_{\bf r}^i {\bf V}^j ,
\end{equation}
where \mbox{$i,j =x,y$} are space indices,~\mbox{${\bf v_p} = \nabla_{\bf p} \epsilon_{\bf p}$} is the quasiparticle velocity, and ${\bf V}$ is a velocity field with \mbox{$\nabla_{\bf r} \cdot {\bf V}=0$}. We impose a harmonic form of the velocity field with momentum ${\bf q}$ and frequency $\omega$, i.e., \mbox{${\bf V} = {\bf V_0} e^{-i \omega t+i {\bf q \cdot r}}$} with ${\bf V_0}$ a constant vector satisfying \mbox{${\bf q \cdot V_0} = 0$}.

The right-hand side of Eq.~(\ref{eq:FL_Boltzmann}) is the collision integral, which describes quasiparticle relaxation due to elastic two-particle collisions,
\begin{align}
    \mathcal{I}&[n_{\bf p}] = \frac{1}{\hbar} \int \frac{d{\bf q}}{(2\pi)^2}\frac{d{\bf p'}}{(2\pi)^2}\frac{d{\bf q'}}{(2\pi)^2} \left|\langle {\bf p',q'} | T | {\bf p,q} \rangle \right|^2 
    \nonumber \\
    & \times  (2\pi)^2\delta({\bf p + q -p'-q'})  2\pi\delta\left(\epsilon_{\bf p} +\epsilon_{\bf q} - \epsilon_{\bf p'} -\epsilon_{\bf q'}\right) \nonumber \\
    & \times  \bigl[\left(1-n({\bf p})\right) \left(1-n({\bf q})\right) n({\bf p'})n({\bf q'}) \nonumber \\
    &\qquad - (\lbrace {\bf p, q}\rbrace \rightarrow \lbrace {\bf p', q'}\rbrace)\bigr] ,
    \label{eq:Icol}
\end{align}
where~\mbox{$\langle {\bf p',q'} | T | {\bf p,q} \rangle$} is the scattering $T$-matrix between states ${\bf p,q}$ and \mbox{${\bf p', q'}$}. The linearized collision integral is then \mbox{$\mathcal{L}[\psi({\bf p})] = \mathcal{I}[n_{\bf p}]/[n_{\rm FD}({\bf p})(1-n_{\rm FD}({\bf p}))]$} with the parameterization of Eq.~\eqref{eq:deltan}. Within Fermi liquid theory, the $T$-matrix is linked to the Landau parameters~$F_m$~\cite{baym04}: 
In terms of the dominant Landau parameter $F_0$ and at low temperatures, 
it is a function of the scaling variable $x = \omega/(v_F|{\bf q-p'}|)$, 
\begin{align}
    \langle {\bf p',q'} | T | {\bf p,q} \rangle &= \frac{1}{\nu}\frac{F_0}{1+F_0 \Omega(x)} ,
\end{align}
with the in-medium dressing function 
\begin{align}
    \Omega(x) &\approx \biggl[1- \frac{x}{\sqrt{(x+i\delta)^2-1}}\biggr] \theta(\mu) .
    \label{eq:t_matrix}
\end{align}
This function parametrizes repeated particle-hole scattering and is equivalent to the Lindhard function in 2D. Equation~\eqref{eq:t_matrix} holds for a well-defined Fermi surface with \mbox{$\mu >0$}. In the opposite limit \mbox{$\mu<0$} (i.e., the BEC or the high-temperature limit), other processes like particle-particle scattering will be more relevant than the ones captured by Fermi liquid theory. For good measure, Supplementary Note I compares and contrasts other forms of the scattering $T$-matrix, which give quantitatively similar results.
The dominant Landau parameter~$F_0$ is connected to the inverse compressibility as~\mbox{$\kappa^{-1} = (1+F_0) \kappa_0^{-1}$} with~$\kappa_0^{-1}$ the inverse compressibility of a non-interacting Fermi gas. This compressibility has been previously measured experimentally over the entire BCS-BEC crossover~\cite{Fenech2016, Boettcher2016,Makhalov2014}, which gives the Landau parameter $F_0$ in terms of the interaction parameter $\ln(k_Fa_2)$. The fit is discussed in the Supplementary Note III.

We solve the kinetic equation by expanding the quasiparticle distribution function to first order in the deviation from local equilibrium, \mbox{$\delta n({\bf p})$}, as well as the slowly varying velocity field ${\bf V}$. By employing a temperature-dependent eigenfunction expansion developed in~\cite{hofmann23,nilsson24}, we obtain the full spectrum of eigenmodes of the collision integral \mbox{$\left\lbrace 1/\tau_m \right\rbrace$}, which describe the relaxation rates of different harmonic modes. In addition, this procedure allows for an exact solution of the linearized Boltzmann equation \mbox{$\delta n({\bf p})$} in response to the shear flow ${\bf V}$. The solution of \mbox{$\delta n({\bf p})$} then gives the linear response of the off-diagonal elements of the stress tensor,
\begin{align}
    \Pi^{xy} &= \int \frac{d{\bf p}}{(2\pi)^2} m^* v_{p_x} v_{p_y} \delta n({\bf p}) \approx i \eta(\omega, q) q_x V^0_y ,
    \label{eq:Pi_def}
\end{align}
from which we compute the shear viscosity spectral function \mbox{$\eta(\omega,q)$}.

{\it \bf Acknowledgments} 
JM is partially supported by the Provincia Autonoma di Trento. This work is supported by Vetenskapsr\aa det (Grant Nos.~2020-04239 and 2024-04485), the Olle Engkvist Foundation  (Grant No.~233-0339), the Knut and Alice Wallenberg Foundation (Grant No. KAW 2024.0129), and Nordita. The computations were enabled by resources provided by the National Academic Infrastructure for Supercomputing in Sweden (NAISS) at Linköping University partially funded by the Swedish Research Council through grant agreement no. 2022-06725.

{\it \bf Author contributions} 
JM, UG, and JH all contributed to equally to the conception, computations, and preparation of the manuscript. 

{\it \bf Competing interests} The authors declare no competing interests.

{\it \bf Correspondence} should be addressed to any of the authors.

{\it \bf Code availability} requests should be addressed to any of the authors.

{\it \bf Data availability}
The raw data for the figures are available upon request addressed to any of the authors.

\bibliography{bib_atomic_gas} 

\end{document}


\title{Supplementary Information: \\
Odd-parity effect and scale-dependent viscosity in atomic quantum gases} 

\author{Jeff Maki}
\affiliation{Department of Physics, University of Konstanz, 78464 Konstanz, Germany}
\affiliation{Pitaevskii BEC Center, CNR-INO and Dipartimento di Fisica, Universit\`a di Trento, I-38123 Trento, Italy}

\author{Ulf Gran}
\affiliation{Department of Physics, Chalmers University of Technology, 41296 Gothenburg, Sweden}

\author{Johannes Hofmann}
\affiliation{Department of Physics, Gothenburg University, 41296 Gothenburg, Sweden}
\affiliation{Nordita, Stockholm University and KTH Royal Institute of Technology, 10691 Stockholm, Sweden}

\date{\today}

\maketitle

This Supplementary Information contains details on the evaluation of the Fermi-liquid collision integral including different $T$-matrix approximations, the computation of the stress response and shear viscosity, and a fit of the Landau parameter $F_0$ that sets the quasiparticle interaction strength.

\section*{Supplementary Note 1: Fermi-liquid collision integral}

The rate of quasiparticle relaxation due to elastic two-particle scattering is given by the collision integral, which takes the form
\begin{align}
    \mathcal{I}&[n_{\bf p}] = \frac{1}{\hbar}\int \frac{d{\bf q}}{(2\pi)^2}\frac{d{\bf p'}}{(2\pi)^2}\frac{d{\bf q'}}{(2\pi)^2} W({\bf p,q; p',q'})  \nonumber \\
    &\left[\left(1-n({\bf p})\right) \left(1-n({\bf q})\right) n({\bf p'})n({\bf q'}) -\lbrace {\bf p, q}\rbrace \rightarrow \lbrace {\bf p', q'}\rbrace\right] . 
\end{align}
In the above equation, \mbox{$n({\bf p}) = n(t,{\bf r , p})$} is the quasiparticle distribution at momentum \mbox{${\bf p}$}, time \mbox{$t$}, and position \mbox{$\bf r$}. For simplicity, we omit  space-time coordinates in our notation. The above expression for the collision integral contains the transition element
\begin{align}
    & W({\bf p',q';p,q}) = \left|\langle {\bf p',q'} | T | {\bf p,q} \rangle \right|^2\nonumber \\
    & \times  (2\pi)^2\delta({\bf p + q -p'-q'})  2\pi\delta\left(\epsilon_{\bf p} +\epsilon_{\bf q} - \epsilon_{\bf p'} -\epsilon_{\bf q'}\right) , 
\end{align}
which depends on the two-body scattering $T$-matrix with initial wave vectors ${\bf p,q}$ and final wave vectors ${\bf p', q'}$. In our calculations, we examine three natural expressions for this quantity: (i) a constant matrix element, (ii) two-body scattering in a medium (sometimes referred to as the many-body $T$-matrix), and (iii) repeated particle-hole scattering. As anticipated in the main text for the odd-even effect in the lifetimes, which originates from phase-space restrictions imposed by Pauli blocking, all of these approaches predict an odd-even effect in the relaxation rates of odd and even modes for sufficiently large \mbox{$\beta \mu$} (as shown below, cf.~Fig.~\ref{fig:eigenvalues_comparison})~\cite{hofmann23}, with quantitative improvements for in-medium and particle-hole corrections over the constant matrix element results. 

For two-component Fermi gases, two-body scattering is dominated by an isotropic contact interaction of strength \mbox{$\hbar^2\tilde{g}/m^*$} with \mbox{$\tilde{g}= -2\pi / \ln(k_F a_2)$}, where $k_F$ is the Fermi wavenumber and $a_2$ is the 2D scattering length~\cite{bloch08}. Such an interaction is captured by the isotropic Landau parameter $F_0$, which itself is a function of~$\tilde{g}$ that can be extracted from experimental data of the equation of state (see the third Supplementary section).

In detail, the first approximation assumes no kinematic dependence of the scattering matrix element on the scattering wavevectors, which is replaced by a constant,
\begin{align}
\langle {\bf p', q'} | T | {\bf p, q}\rangle &= \frac{1}{\nu}T({\bf Q}, {\bf k}),
\end{align} 
with
\begin{align}
    T({\bf Q}, {\bf k}) &=  F_0, \label{eq:Tconst}
\end{align} 
where $F_0$ is the $s$-wave Fermi-liquid parameter, \mbox{$\nu = m^*/(\pi \hbar^2)$} is the density of states, and $k_F$ is the Fermi wavenumber.

\begin{figure}
    \centering
    \includegraphics{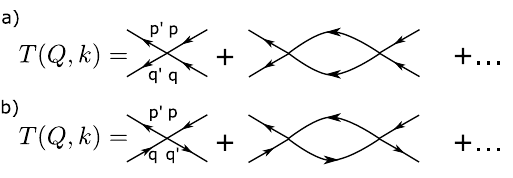}
    \caption{Diagrams contributing to the two-body $T$-matrix. (a) Standard approach for cold atoms that accounts for repeated particle-particle scattering. (b) In the Fermi liquid picture, the quasiparticles are dressed by repeated particle-hole scattering. }
    \label{fig:diagrams}
\end{figure}

For the second approximation, the two-body in-medium scattering $T$-matrix sums the diagrams shown in Fig.~\ref{fig:diagrams}(a), which gives
\begin{align}
T^{-1}({\bf Q},{\bf k}) &= \frac{1}{4} \ln\left(\frac{1}{(k a_2)^2}\right) \nonumber \\
&+ \int \frac{d{\bf l}}{2\pi} \frac{n_{\rm FD}\left(\bf l\right)}{\frac{Q^2}{4}+k^2-l^2 - {\bf Q \cdot l}}~,
\label{eq:T}
\end{align}
where \mbox{${\bf Q = p+q=p'+q'}$} is the total wavevector, \mbox{${\bf k} = {\bf (p-q)}/2$} and \mbox{${\bf k' = (p'-q')}/2$} are the relative wavevectors before and after the collision,~\mbox{$a_2$} is the two-dimensional scattering length, and~\mbox{$n_{\rm FD}({\bf k}) = (e^{\beta (\hbar^2 k^2/2m^*-\mu)}+1)^{-1}$} is the Fermi-Dirac distribution. In Eq.~\eqref{eq:T}, we place the $T$-matrix on-shell, i.e., the frequency is set to the energy of the scattering atoms. This common approximation becomes exact at high-temperatures~\cite{barth14}, where it captures the leading Pauli-blocking effect in a nondegenerate gas. However, one could expect that this approach becomes inaccurate at the low temperatures that are the focus of our study.

For this reason, results in the main text are presented using the third scheme, which is the canonical Fermi-liquid form valid at low temperature, and for which we use experimental data as input for the Landau parameters. Here, the two-body $T$-matrix sums repeated particle-hole scattering as shown in Fig.~\ref{fig:diagrams}(b). The sum of the diagrams is equivalent to the Bethe-Salpeter equation
\begin{align}
T&({\bf Q,k}) = F_{\bf \frac{Q}{2}-k,\frac{Q}{2}-k'} + \frac{1}{\nu}\int \frac{d^2{\bf l}}{(2\pi)^2} F_{\bf \frac{Q}{2}-k, l} T(\bf l, \frac{Q}{2}-k') \nonumber \\
&\times  \frac{n_{\rm FD}\left({\bf l} +\frac{\bf (k'+k)}{2}\right)-n_{\rm FD}\left({\bf l} -\frac{\bf (k'+k)}{2}\right)}{2\left( {\bf Q \cdot k - l \cdot(k+k')}\right)} ,
\label{eq:t_matrix_FL}
\end{align}
where the bare vertex is the Landau \mbox{$F_{\bf p,p'}$} parameters [cf.~Eq.~(4) of the main text], we  put the frequency of the $T$-matrix on shell, and we abbreviate \mbox{$\xi_{\bf k} = \hbar^2 k^2/(2m^*)-\mu$} \cite{baym04}. The above expression can be simplified further for $s$-wave interactions where~\mbox{$F_{\bf p,p'}$} is dominated by the zero-angular momentum mode~\mbox{$F_0$}. In this case, we can drop all angular dependence on~\mbox{${\bf k}$} and~\mbox{${\bf k'}$} when solving Eq.~\eqref{eq:t_matrix_FL}.

To compare these three approaches, Fig.~\ref{fig:eigenvalues_comparison} shows the lowest~\mbox{$m=2$} (continuous lines, even parity) and~\mbox{$m=3$} (dashed lines, odd parity) eigenvalues for the linearized collision integrals as obtained using a constant matrix element~\eqref{eq:Tconst} (blue line), the many-body $T$-matrix~\eqref{eq:T} (orange line), and Fermi liquid theory~\eqref{eq:t_matrix_FL} (green line). We use \mbox{$F_0 = 0.2$} for the Fermi-liquid calculation (blue),
\mbox{$\ln(k_F a_2) = 0$} for the many-body $T$-matrix (orange), and a constant matrix element that is set to unity (green). In the figure, we have rescaled the relaxation rates by a constant value to highlight the qualitative similarities of all approximations. As anticipated above, in all cases the results differ quantitatively but not qualitatively. In the BCS limit \mbox{$(\beta \mu>0)$}, there is a clear separation of the odd- and even-modes, while near resonance and in the BEC limit \mbox{$(\beta\mu <0)$} the odd- and even-modes are of the same order. This illustrates that the odd-even effect is very general and relies on the restricted phase space of quasi-particle scattering due to the concomitant Pauli blocking from the Fermi surface.

\begin{figure}
    \centering
    \includegraphics{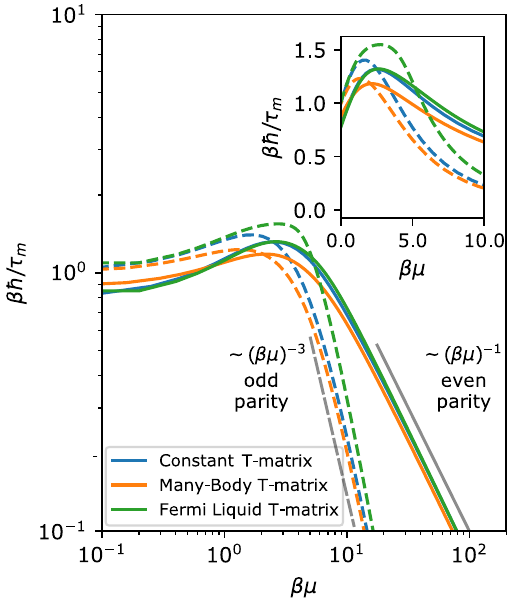}
    \caption{Comparison of the leading odd and even relaxation rates as calculated from the linearized collision integral, Eq.~\eqref{eq:linearized_collision_integral}, where the transition element has been calculated via a constant matrix element with \mbox{$F_0=1$} [Eq.~\eqref{eq:Tconst}, blue line], the many-body $T$-matrix with resonant particle-particle scattering \mbox{$\ln(k_F a) = 0$} [Eq.~\eqref{eq:T}, orange line], and the Fermi liquid $T$-matrix with~\mbox{$F_0 = 0.2$} [Eq.~\eqref{eq:t_matrix_FL}, green line]. The solid (dashed) lines show the longest-lived even-parity~\mbox{$m=2$} and odd-parity~\mbox{$m=3$} relaxation modes, respectively. Gray lines show the asymptotic scaling. The inset shows the same relaxation rates on a linear scale for comparison. All three $T$-matrix expressions predict the odd-even effect at large values of $\beta\mu$.}
    \label{fig:eigenvalues_comparison}
\end{figure}

\section*{Supplementary Note 2: Stress response and shear viscosity}

We solve the Fermi-liquid kinetic equation for small deviations in the fermion distribution function induced by the external velocity field ${\bf V}$. 
To linear order in $\psi({\bf p})$, the Fermi-liquid kinetic equation reads:
\begin{align}
&-i\mathcal{L}\left[\psi({\bf p})\right] =\nonumber \\
&\quad \left[\omega -{\bf v_p \cdot q}\right] \psi({\bf p})- \beta m^*\left({\bf v_p \cdot q}\right) \left({\bf v_p \cdot V_0}\right),
\label{eq:final_linearized_Boltzmann}
\end{align}
where the deviation function $\psi$ is defined in Eq.~(1) of the main text and we define the linearized collision integral:
\begin{align}
    \mathcal{L}[\psi({\bf p})] &= \frac{1}{\hbar}\int \frac{d{\bf q}}{(2\pi)^2}\frac{d{\bf p'}}{(2\pi)^2}\frac{d{\bf q'}}{(2\pi)^2} W({\bf p,q; p',q'}) \nonumber \\
    &\times \frac{n_{\rm FD}({\bf q})\left(1-n_{\rm FD}({\bf p'})\right)\left(1-n_{\rm FD}({\bf q'})\right)}{\left(1-n_{\rm FD}({\bf p})\right)} \nonumber \\
    &\times \left[\psi({\bf p}) + \psi({\bf q})- \psi({\bf p'})-\psi({\bf q'})\right] .
    \label{eq:linearized_collision_integral}
\end{align}
We solve Eq.~\eqref{eq:final_linearized_Boltzmann} using a method developed by two of the authors~\cite{hofmann23,nilsson24} for two-dimensional electron gases. To this end, we expand the deviation function~$\psi({\bf p})$ in terms of a temperature-dependent  set of  polynomials that are orthogonal with respect to the inner-product:
\begin{equation}
\langle g| h \rangle = \lambda_T^2\int \frac{d{\bf p}}{(2\pi)^2} n_{\rm FD}({\bf p}) \left(1-n_{\rm FD}({\bf p}) \right) g({\bf p}) h({\bf p})~.
\label{eq:inner_product}
\end{equation}
These orthogonal modes, which we denote as $\chi_{n,m}(w)$, are functions of the variable \mbox{$w = \beta(\hbar^2 p^2/(2m^*)-\mu)$}, and are labeled by a principal quantum number $n$ and an angular quantum number $m$. Expanding $\psi({\bf p})$ in Eq.~\eqref{eq:final_linearized_Boltzmann} in terms of these modes,
\begin{equation}
    \psi({\bf p}) = \sum_{n,m} c_{n,m} \chi_{n,m}(w) e^{i m \theta_{\bf p}},
\end{equation}
and evaluating the matrix elements of the streaming term and collision integral with respect to these basis functions gives a linear system of equations for the expansion coefficients~$c_{n,m}$. Inverting Eq.~\eqref{eq:final_linearized_Boltzmann} for $c_{n,m}$ then provides the full solution for $\delta n_{\bf p}$ in response to the external velocity field.

Using the full solution of the deformation function, we obtain the off-diagonal element of the stress-tensor:
\begin{align}
    &\Pi^{xy} = \int \frac{d{\bf p}}{(2\pi)^2} m^*{\bf v_p}^x {\bf v_p}^y\delta n_{\bf p} \nonumber \\
    &= \sum_{n,m}\frac{i}{4} \bigl(\delta_{m,2}-\delta_{m,-2}\bigr) c_{n,m} \nonumber \\
    &\times \int \frac{d{\bf p}}{(2\pi)^2} n_{\rm FD}({\bf p})(1-n_{\rm FD}({\bf p}))m^* {\bf v_p}^2 \chi_{n,m}(w) .
    \label{eq:Pi_def}
\end{align}
In the limit of small transverse velocity fields, the stress-tensor is proportional to $i{\bf q}^x {\bf V}^y$, with the constant of proportionality defining the shear viscosity spectral function $\eta(\omega,{\bf q})$,
\begin{equation}
    \Pi^{x,y}\approx i \eta(\omega,{\bf q}) {\bf q}^x {\bf V_0}^y~.
    \label{eq:shear_def}
\end{equation}
The Boltzmann equation thus gives access to the shear viscosity spectral function at arbitrary values of the wavevector ${\bf q}$ and frequency $\omega$.

\begin{figure}[b!]
    \centering\includegraphics{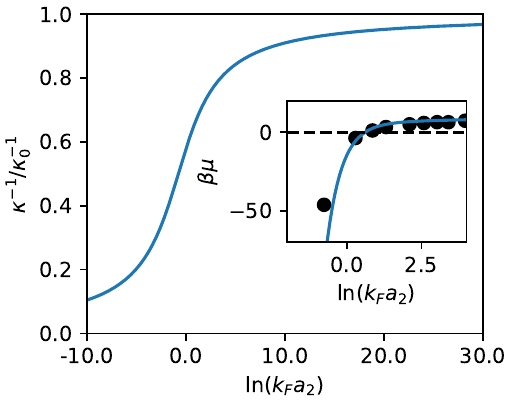}
\caption{Compressibility extracted using the ansatz~\eqref{eq:fiteos} from the equation of state of a 2D Fermi gas along the BEC-BCS crossover at a fixed low temperature \mbox{$T/T_F = 0.1$}, shown here as a function of the dimensionless interaction strength $\ln (k_Fa_2)$. The compressibility is related to the Fermi liquid parameter $F_0^s$ as \mbox{$\kappa^{-1}/\kappa_0^{-1} \approx 1+F_0^s$}, where \mbox{$\kappa_0^{-1} = \pi \hbar^2 n^2/(2m^*)$} is the  compressibility of the noninteracting gas. The inset shows the experimental data for the chemical potential from Ref.~\cite{Boettcher2016}, while the solid line is the interpolation in Eqs.~(\ref{eq:fiteos}-\ref{eq:c_phen}).}
    \label{fig:EoS}
\end{figure}

\section*{Supplementary Note 3: Equation of state}
\label{app:eos}

Within Fermi liquid theory, the isotropic Fermi-liquid parameter $F_0$ determines the inverse compressibility of the system, $\kappa^{-1} = n^2 \partial \mu /\partial n$, as
\begin{equation}
    \frac{\kappa_0}{\kappa} \approx 1+F_0 ,
\end{equation}
where \mbox{$\kappa_0^{-1} = \pi \hbar^2 n^2/(2m^*)$} is the bare inverse compressibility of a gas of quasiparticles with mass~$m^*$, where we neglect the mass renormalization. The compressibility (and thus the Fermi liquid parameter for a given interaction strength $\ln(k_Fa_2)$) then follows from the equation of state, which has been extensively studied both theoretically \cite{Bertaina2011,Mulkerin2015,bauer14,Anderson2015} and experimentally \cite{Fenech2016, Boettcher2016,Makhalov2014}. Here, we use the experimental data from Ref.~\cite{Boettcher2016} to extract $F_0$, with results for the chemical potential shown in the inset of Fig.~\ref{fig:EoS}.

Following Refs.~\cite{Makhalov2014, Boettcher2016}, we parametrize the data using
\begin{equation}
    \mu = -\frac{\epsilon_b}{2} + \epsilon_F c(\ln(k_F a_2)) , \label{eq:fiteos}
\end{equation}
where \mbox{$\epsilon_b = \hbar^2/(m^*a_2^2)$} is the two-body binding energy and the function $c(x)$ parametrizes the interaction dependence of the chemical potential. We approximate this function $c(x)$ using the following form:
\begin{equation}
c(x) = \frac{1}{\pi}\arctan\left(\frac{1}{\pi}\left(x + (1-\ln(2)\right)\right) + \frac{1}{2}~.
\label{eq:c_phen}
\end{equation}
The coefficients in Eq.~\eqref{eq:c_phen} are chosen such that they reproduce the perturbative results from Fermi liquid theory~\cite{Engelbrecht1992},
\begin{align}
F_0 &= 1 - \frac{1}{\ln(k_F a_2)} + {\it O}\biggl(\frac{1}{\ln(k_F a_2)^2}\biggr) .
\end{align}
It is then straightforward to express the compressibility in terms of $c(x)$ and to obtain the Landau parameter as
\begin{align}
    \frac{\kappa^{-1}}{\kappa_0^{-1}} &\approx 1+F_0^s = c(\ln(k_Fa_2)) + \frac{1}{2}c'(\ln(k_Fa_2))~.
\end{align}
This result is shown in Fig.~\ref{fig:EoS} as a blue continuous line. 
Equation~\eqref{eq:c_phen} also qualitatively captures the experimental data for the chemical potential along the crossover, see the inset of Fig.~\ref{fig:EoS} for a comparison. 

\bibliography{bib_atomic_gas}